# The Naturalness of the Fourth SM Family

Saleh Sultansoy

TOBB ETU, Ankara, Turkey
Institute of Physics, Baku, Azerbaijan

**Abstract.** - The necessity of the fourth family follows from the SM basics. According to flavor democracy the Dirac masses of the fourth SM family fermions are almost equal with preferable value 450 GeV, which corresponds to common (for all fundamental fermions) Yukawa coupling equal to SU(2) gauge coupling $g_W$. In principle, one expect $u_4$ a little bit lighter than $d_4$, while $v_4$ could be essentially lighter than $l_4$ due to Majorana mass terms for right-handed components of neutrinos. Obviously, the fourth family quarks will be copiously produced at the LHC. However, the first indication of the fourth SM family may be provided by early Higgs boson observation due to almost an order enhancement of the gluon fusion to Higgs cross-section. For the same reason the Tevatron still has a chance to observe the Higgs boson before the LHC. Concerning the fourth family leptons, in general, best place will be NLC/CLIC. However, for some mass regions and MNS matrix elements double discovery of both the $v_4$ and H could be possible at the LHC.

## *Prologue*

*30 November 2007: 57 people were killed in airplane crash near Isparta, Turkey. Turkish Physics Community loses 6 brilliant members, participants of the Turkic Accelerator Complex (TAC) Project. All of us were shocked, but my personal tragedy was deeper, because I lose my big sister Engin Arık and my brother Engin Abat (one of the most promising young scientists in our field). Professor Engin Arık was bannerbearer of the High Energy Physics in Turkey and, moreover, in Turkic World. Her dreams were: CERN membership for Turkey, TAC for developing of Turkey and neighbor countries, sustainable development for Mankind ... Her scientific interests cover almost whole complex of HEP connected fields (see, for example [1]). Concerning the fourth SM family, an essential part of Turkish group activity (especially Higgs boson related topics) was initiated by Engin Arık and performed by her students.*

*I apologize that this paper is not sufficiently perfect to be worthy of memory of my colleagues (I hope to finish extended version of the paper during this year).*

## 1. A little bit history

Particle physics in 1930's: electron, photon, proton and neutron are discovered, neutrino and π-mesons are predicted by Pauli and Yukawa, respectively. EM interactions are mediated by photons, strong interactions are thought mediated by π-mesons and weak interactions are described through four-fermion contact interaction as proposed by Fermi. There is a "simple" classification of matter fields: leptons (e and ν), mesons ($π^±$ and $π^0$), barions (p and n). Whole (visible) Universe is formed from a few particles: nuclei are bound states of p's and n's, atoms are bound states of nuclei and e's etc. Chemistry became the Science… Whole technology of 20th century is based on this picture.

**This nice picture was destroyed in 1937 by the discovery of μ!** We were looked for π–mesons but found something different. This new particle seems to be produced by strong interactions, but interacts with matter by EM interactions. Real π–mesons were discovered 10 years later in emulsion experiments… We fall with **μ–e puzzle:** why the Nature needs the second "heavy" electron …

1960's: hadron (meson and barion) inflation leads to the Quark Model.

1970's: GİM $\Rightarrow$ c-quark (also from *q-l* symmetry as the counterpart of $v_\mu$). Modified **μ–e puzzle:** why the Nature needs 2 families. Experiment: charmed hadrons + τ-lepton + beauty. Theory: CKM $\Rightarrow$ 3 families (CP phase, BAU).

1990's: Experiment: t-quark, $m_H > 114$ GeV. Theory: fourth SM family revisited (and a lot of different BSM models proposed during last decades).

Let me continue with an introductory paragraph from 10 years old paper [2]: "It is known, that the Standard Model with three fermion families well describes almost all of the large amount of particle physics phenomena. Today, SM is proved at the level of first-order radiative corrections for energies up to 100 GeV. However, there are a number of fundamental problems which do not have solutions in the framework of the SM: quark-lepton symmetry and fermion's mass and mixing pattern, family replication and number of families, L-R symmetry breaking, electroweak scale etc. Then, SM contains unacceptably large number of arbitrary parameters even in three family case: 19 in the absence of right neutrinos (and Majorana mass terms for left neutrinos), 26 if neutrinos are Dirac particles, > 30 if neutrinos are Majorana particles. Moreover, the number of "elementary particles", which is equal to 37 in three families case (18 quarks, 6 leptons, 12 gauge bosons and 1 Higgs boson), reminds the Mendeleyev Table. Three decades ago similar situation led to the quark model!"

Flavor Democracy hypothesis provides an opportunity to solve some of mentioned problems and predicts the fourth SM family (see review [3] and references therein). The fourth family matters were discussed in deep during the topical workshop [4] held in September 2008 at CERN (see [5] for resume of the workshop).

## 2. Why the Four SM Families?

There are two possible answers to this question. The first one is "Why not?" Indeed, lower bound on the number of families comes from LEP data (N ≥ 3), and upper bound is 8 from asymptotic freedom. The second answer (preferred by our group) is: existence of the fourth SM family follows from the SM basics and phenomenology (Flavor Democracy).

There were two (incorrect/wrong) objections: incorrect interpretation of LEP1 data and wrong interpretation of EW precision data. Concerning the first one: only the number of "active" (in SM LH ν), "massless" neutrinos (with $2m < m_Z$) is equal to 3. In some sense it is the consequence of the historical "paralogism" (V-A → ν ≡ $v_L$), but according the SM (q-l symmetry) RH ν should exist, because it is the partner of RH up-quark. Concerning the second one, namely "A 4th generation of ordinary fermions is excluded to 99.999% CL on the basis of S parameter alone" (see, e g PDG 2006), the clearest answer, namely, "This conclusion is wrong**.**" was given by G. Kribbs in August 2007 at CERN seminar (see, also, [6-8]).

**The clear arguments for fourth SM family** are given in original papers [9-11] and reviews [2, 3, 12]. In the case of entire democracy the (Dirac) masses of the fourth SM family fermions are almost degenerate with common value between 300 GeV and 700 GeV, whereas all known fermions are massless. Fermions from first three generations acquire masses as a result of slight violation of full democracy. A number of parameterizations of slightly broken Democratic Mass Matrix, which give opportunity to obtain both mass and mixing patterns with a fewer number of parameters are proposed [13-15]. Last parameterization, which gives correct values for fundamental fermion masses, at the same time, predicts quark and lepton CKM matrices in good agreement with experimental data.

Flavor Democracy essentially reduces the number of free parameters, which is equal to 26 in SM3 and 40 in SM4 for the Dirac neutrino case. For example, with parameterization [15] last number fall from 40 to 25. My personal feeling: Dirac masses of the fourth SM family fermions are almost equal with preferable value 450 GeV (500-550 for the pole mass), which corresponds to common (for all fundamental fermions) Yukawa coupling equal to SU(2) gauge coupling $g_W$; $u_4$ a little bit lighter than $d_4$, while $\nu_4$ could be essentially lighter than $l_4$ due to Majorana mass terms for right-handed components of neutrinos. It is interest, that if the quartic coupling constant of the Higgs self-interaction is also equal to $g_W$, the Higgs boson mass is predicted to be around 300 GeV.

*Arguments Against the Fifth SM Family:* The first argument disfavoring the fifth SM family is the large value of $m_t \approx 175$ GeV. Indeed, partial-wave unitarity leads to $m_Q \leq 700$ GeV $\approx 4m_t$ and in general we expect $m_t \ll m_4 \ll m_5$. Then, neutrino counting at LEP results in fact that there are only three "light" ($2m_\nu < m_Z$) non-sterile neutrinos, whereas in the case of five SM families four "light" neutrinos are expected. Finally, the fifth family is excluded at more than 5σ level by precision electroweak data.

Below I give some remarks on search for fourth SM family including several new results obtained since 2006 (mainly by Turkish group). For more adequate picture see [4, 5].

## 3. The Fourth SM Family at Tevatron/LHC

Hadron colliders are best place for discovery of new quarks. Together with direct production, fourth family quarks will manifest themselves due to an essential enhancement of the Higgs boson production via gluon fusion.

### 3.1. Direct Search

The fourth SM family quarks will be copiously produced at the LHC via gluon-gluon fusion (see [16] and references therein). The expected cross section is about 10 (0.25) pb for a quark mass of 400 (800) GeV. The fourth generation up-type quark, $u_4$, would predominantly decay via $u_4 \rightarrow Wb$, therefore, the expected event topologies are similar to those for t-quark pair production. The best channel for observing will be [17]:

$$gg \rightarrow u_4 \bar{u}_4 \rightarrow WWb\bar{b} \rightarrow (l\nu)(JJ)b\bar{b}$$

where one W decays leptonically and the other hadronically. The mass resolution is estimated to be 20 (40) GeV for $m_{u4} = 300$ (700) GeV. The small interfamily mixings [14, 15] lead to the formation of the fourth family quarkonia [16, 18]. The most promising candidate for the LHC is the pseudo-scalar quarkonium state, $\eta_4$, which will be produced resonantly via gluon-gluon fusion. Especially, the decay channel $\eta_4 \rightarrow ZH$ is the matter of interest [19].

If the fourth generation quarks prefer to mix with the first two generations, the signal is doubled because $u_4$ and $d_4$ are indistinguishable (mass difference is less than 50 GeV). In this case 500 GeV fourth family quarks can be discovered at the LHC with 400 pb[-1] data [20].

If $\nu_4$ has Majorana nature, for definite values of $\nu_4$ and Higgs boson masses double discovery of both of them can be possible at the LHC with a few fb[-1] data [21].

The FNAL Tevatron Run II can observe $u_4$ and $d_4$ quarks if there is an anomalous interaction with enough strength between the fourth family quarks and known quarks [22]. Recent CDF data on the search for "heavy top" [23] could be interpreted as a hint for fourth family quarks [24].

### 3.2. Higgs Boson

The fourth SM family leads to an essential increase (~8 times) of the Higgs boson production cross-section at hadron colliders and this can give the Tevatron experiments (CDF and D0) opportunity to discover the Higgs boson before the LHC, if its mass is between 140 and 200 GeV (see [25] and ref's therein). Both D0 and CDF Collaborations are looking for this opportunity [26, 27]. Already in 2006, their results were placing constraints on the SM with four or more fermion families [28]. Recently, CDF and D0 combined results exclude the Higgs mass between 160 and 170 GeV for the SM3 [29]. For the SM4 the same results exclude the region approximately between 130 and 220 GeV (approximately, because in the analysis other channels are added to the main/dominate one, namely, gg→H→WW, and region above 200 GeV is not covered).

Concerning the LHC, it will be able to cover the whole region via the golden mode $H \to ZZ \to \ell\ell\ell\ell$ and detect the Higgs signal during the first year of operation if the fourth SM family exists [30-33].

## 4. The Fourth SM Family at ILC/CLIC and QCD-E/LHeC

The fourth family leptons will clearly manifest themselves at the future lepton and photon colliders [15, 34, 35]. Also, the number of different fourth family quarkonium states can be produced resonantly at lepton and photon machines. Moreover, in difference from the LHC, states formed by up and down type quarks can be investigated separately even if their mass difference is small.

Concerning future ep and γp colliders, they have promising potential for search possible anomalous interactions of both fourth family leptons and quarks [36-38].

*Epilogue*

*16 Mart 2009: Turkey sent official letter to the CERN Council in order to start the membership procedure.*

**Acknowledgements:** I am grateful to Serkant Ali ÇETİN and Gökhan ÜNEL for intangible support, as well as to all my colleagues/friends for fruitful collaboration.

**REFERENCES**

[1] E. Arik and S. Sultansoy, *Turkish comments on 'Future perspectives in HEP'*, BOUN-HEP-2003-01, GU-HEP-2003-01, e-Print: hep-ph/0302012 (partially presented at 2002 ICFA Seminar on Future Perspectives in High-Energy Physics, 8-11 October 2002, Geneva, Switzerland).

[2] S. Sultansoy, *Four Ways to TeV Scale*, Proceedings of the International Workshop on Linac-Ring Type ep and γp Colliders, 9-11 April 1997, Ankara, Turkey; published in Tr. J. of Physics **22**, 575, 1998.

[3] S. Sultansoy, *Flavor Democracy in Particle Physics*, Invited talk at 6th International Conference of the Balkan Physical Union (Istanbul, Turkey, 22-26 Aug 2006). Published in AIP Conf. Proc. **899**, 49, 2007. e-Print: hep-ph/0610279

[4] http://indico.cern.ch/conferenceDisplay.py?confId=33285

[5] B. Holdom et al., *Four Statements about the Fourth Generation*, e-Print: arXiv:0904.4698 [hep-ph]

[6] H.J. He, N. Polonsky and S. Su, Phys. Rev. **D 64** (2001) 117701

[7] V.A. Novikov, L.B. Okun, A.N. Rosanov and M.I. Vysotsky, Phys. Lett. **B 529** (2002) 111.

[8] G.D. Kribs, T. Plehn, M. Spannowsky and T.M.P. Tait, Phys. Rev. **D 76** (2007) 075016.

[9] H. Ftitzsch, Phys. Lett. **B 289**, 92 (1992).

[10] A. Datta, Pramana **40**, L503 (1993).

[11] A. Celikel, A.K. Ciftci and S. Sultansoy, Phys. Lett. **B 342**, 257 (1995).

[12] S. Sultansoy, *Why the four SM families*, e-Print: hep-ph/0004271

[13] A. Datta and S. Raychaudhuri, Phys. Rev. **D 49**, 4762 (1994).

[14] S. Atag et al, Phys. Rev. **D 54**, 5745 (1996).

[15] A. K. Ciftci, R. Ciftci and S. Sultansoy, Phys. Rev. **D 72**, 053006 (2005).

[16] ATLAS Detector and Physics Performance Technical Design Report, CERN/LHCC/99-15 (1999), vol. II, pp. 663-668.

[17] E. Arik et al, Phys. Rev. **D 58**, 117701 (1998).

[18] H. Ciftci and S. Sultansoy, Mod. Phys. Lett. **A 18**, 859 (2003).

[19] E. Arik et al, Phys. Rev. **D 66**, 116006 (2002).

[20] E. Ozcan, S. Sultansoy and G. Unel, Eur. Phys. J. C **57**, 621 (2008).

[21] T. Cuhadar-Donszelmann et al., JHEP 0810, 074 (2008).

[22] E. Arik, O. Cakir and S. Sultansoy, Phys. Rev. **D 67**, 035002 (2003); Europhys. Lett. **62**, 332 (2003); Eur. Phys. J. **C 39**, 499 (2005).

[23] http://www-cdf.fnal.gov/physics/new/top/confNotes/cdf9446_tprime_public_2.8.pdf

[24] E. Ozcan, S. Sultansoy and G. Unel, *Hints from Tevatron, a prelude to what?*, e-Print: arXiv:0808.0285 [hep-ph]

[25] E. Arik et al, Acta Physica Polonica **B 37**, 2839 (2006).

[26] D0 Collaboration, Phys. Rev. Lett. **96**, 011801 (2006).

[27] CDF Collaboration, Phys. Rev. Lett. **97**, 081802 (2006).

[28] A. Meyer, *Search for New Phenomena at the Tevatron and at HERA*, e-Print: hep-ex/0610001 (2006).

[29] *Combined CDF and DZero Upper Limits on Standard Model Higgs-Boson Production with up to 4.2 fb$^{-1}$ of Data*, e-Print: arXiv:0903.4001 [hep-ex]

[30] E. Arik et al, *Enhancement of the Standard Model Higgs Boson Production Cross-section with the Fourth Standard Model Family Quarks*, ATLAS Internal Note ATL-PHYS-98-125 (1998).

[31] E. Arik et al, Eur. Phys. J. **C 26**, 9 (2002).

[32] E. Arik et al, Phys. Rev. **D 66**, 003033 (2002).

[33] E. Arik, S.A. Cetin and S. Sultansoy, *The impact of the fourth SM family on the Higgs observability at the LHC*, Balk. Phys. Lett. **15** N4, 1 (2007); e-Print: arXiv:0708.0241 [hep-ph]

[34] R. Ciftci et al, Turk. J. Phys. **27**, 179 (2003).

[35] E. Accomando et al., *Physics at the CLIC multi-TeV linear collider*, CERN-2004-005; e-Print: hep-ph/0412251

[36] A.K. Ciftci, R. Ciftci, H. Duran Yildiz and S. Sultansoy, Mod. Phys. Lett. **A 23**, 1047 (2008).

[37] A.K. Ciftci, R. Ciftci and S. Sultansoy, Phys. Lett. **B 660**, 534 (2008).

[38] R. Ciftci and A.K. Ciftci, *A Comperative Study of the Anomalous Single Production of the Fourth Generation Quarks at ep and gamma-p Colliders*, e-Print: arXiv:0904.4489 [hep-ph]